\documentstyle[aps,epsfig,prl] {revtex}
\begin{document}
\title{The magnetopolaron effect in light reflection ana absorption by a wide quantum}
\author{I. G. Lang, L. I. Korovin}
\address{A. F. Ioffe Physical-Technical Institute, Russian Academy of Sciences,
194021 St. Petersburg, Russia}
\author{S. T. Pavlov\dag\ddag}
\address{\dag Facultad de Fisica de la UAZ, Apartado Postal C-580, 98060 Zacatecas, Zac., Mexico\\
\ddag P. N. Lebedev Physical Institute, Russian Academy of
Sciences, 119991 Moscow, Russia}
\twocolumn[\hsize\textwidth\columnwidth\hsize\csname
@twocolumnfalse\endcsname
\date{\today} \maketitle \widetext
\begin{abstract}
\begin{center}
\parbox{6in}{Light reflection and absorption spectra by a semiconductor quantum well (QW) ,
which width is comparable to a light wave length of stimulating
radiation, are calculated. A resonance with two close located
exited levels is considered. These levels can arise due to
splitting of an energy level of an electron-hole pair (EHP) due to
magnetopolaron effect, if the QW is in a quantizing magnetic field
directed perpendicularly to the QW plane. It is shown that unlike
a case of narrow QWs light reflection and absorption depend on a
QW width $d$. The theory is applicable at any ratio of radiative
and non-radiative broadenings of electronic excitations.}
\end{center}
\end{abstract}

] \narrowtext

At passage of light through a semiconductor quantum well (QW) in
reflected and transmitted light waves characteristic occur, on
which it is possible to judge features of electronic processes
proceeding in a QW. The most interesting results turn out  when
energy levels of an electronic system are discrete. It takes place
in a quantizing magnetic field directed perpendicularly to a QW
plane or in case of a resonance of the stimulating light frequency
$\omega_\ell$ with an excitonic state in a zero magnetic field.

Two close located energy levels of excitation arise in a case of a
magneto-phonon resonance [1], when the condition
\begin{equation}
\label{l} \omega_{LO}=j\Omega
\end{equation}
is satisfied, where $\omega {LO}$ is the longitudinal optical (LO)
phonon frequency, $\Omega=|e|H/cm_{e(h)}$ is the cyclotron
frequency, $m_{e(h)}$ is the electron (hole) effective mass.
Integers $j$ correspond to ordinary polarons, fractional $j$
correspond to so-called "weak" polarons [2]. Weak polarons arise
in crossing points of two terms characterized by values $\Delta
N\geq 2$, where $\Delta N$ is the difference of  LO phonons
numbers for these two terms. Real direct transitions between such
terms with emitting of one phonon are impossible. But as the terms
are crossed, their splitting is inevitable. To calculate the
splitting value it is necessary to take into account transitions
between crossed terms through virtual intermediate states or to
take into account in the operator of the electron-phonon
interaction small two-phonon contributions. We obtain that
splitting $\Delta E {weak}$ for weak polarons should be
essentially less, than for usual polarons at integer $j$ .
Contributions of transitions through intermediate states in
$\Delta E_{weak}$ are of higher order on the Fr\"ohlich
electron-phonon coupling constant $\alpha$ [3], than
$\alpha^{1/2}.$

Light scattering and absorption by narrow QWs are theoretically
investigated in conditions of a magnetophonon resonance, when a
light wave length is much greater than a QW width $d$, in [4] - at
a pulse irradiation, in [5] - at a monochromatic irradiation. It
was shown that especially interesting results may be obtained for
weak polarons. Monochromatic light reflection, absorption and
transmission are theoretically investigated for wide QW, when the
interaction of light with two close located energy levels (for
instance, two energy levels of an usual or weak polaron) is
essential. A condition of a wide QW is as follows
$$\kappa d\geq1,$$
where $\kappa$ is the wave vector module of a light wave. The
normal incidence of light on a semiconductor QW surface in a plane
$xy$ is considered. The QW may be in a quantizing magnetic field,
perpendicular to the QW surface. For calculation of light
reflection and absorption the technique developed in [6] and
applicable to any low-dimensional semiconductor objects (QWs,
quantum wires and dots) is used at monochromatic and at pulse
irradiation at any light pulse form. For dimensionless light
reflection $\cal R$ and absorption $\cal A$ in case of two close
located excitation energy levels $\hbar\omega_1$ and
$\hbar\omega_2$ following results are obtained:
\begin{eqnarray}
\label{2} {\cal R}={1\over4Z}\{ [\tilde{\gamma}_{r1}
(\omega_\ell-\omega_2) + \tilde{\gamma}_{r2}
(\omega_\ell-\omega_1)]^2\nonumber\\ + {1 \over 4} (\ tilde
{\gamma} - {r1} \gamma_2+ \tilde {\gamma} - {r2} \gamma_1)^2 \}
\end{eqnarray}
\begin{eqnarray}
\label{3} {\cal A}={1\over 2Z}\{\tilde{\gamma}_{r1}\gamma_1
[(\omega_\ell-\omega_2)^2+\gamma_2^2/4]\nonumber\\+
\tilde{\gamma}_{r2}\gamma_2[(\omega_\ell-\omega_1)^2+\gamma_1^2/4]\nonumber\\+
(\Delta_1\tilde{\gamma}_{r2}-\Delta_2\tilde{\gamma}_{r1})
[(\omega_\ell-\omega_2)\gamma_1-(\omega_\ell-\omega_1)\gamma_2]\},
\end{eqnarray}
\begin{eqnarray}
\label{4} Z=\{(\omega_\ell-\omega_1)(\omega_\ell-\omega_2)-
{\tilde{\gamma}_{r1}\gamma_2+\tilde{\gamma}_{r2}\gamma_1+
\gamma_1\gamma_2\over
4}\nonumber\\-\Delta_1(\omega_\ell-\omega_2)-
\Delta_2(\omega_\ell-\omega_1)\}^2\nonumber\\+
 +{1\over
4}\{(\omega_\ell-\omega_2)(\tilde{\gamma}_{r1}+\gamma_1)+
(\omega_\ell-\omega_1)(\tilde{\gamma}_{r2}+\gamma_2)\nonumber\\
-\Delta_1\gamma_2-\Delta_2\gamma_1\}^2,
\end{eqnarray}
where $\tilde{\gamma}_{r1(2)}$ is the radiative broadening of the
states 1(2), $\gamma_{1(2)}$ is the non-radiative broadening of
the same states, $\Delta_{1(2)}$ is the radiative energy shift of
states. The formulas (2) - (4) are fair, if excitations 1 and 2
are characterized by the same wave function dependent on
coordinates $z$. The last is carried out for usual polarons (but
not for combined polarons [7,8]). For wide QWs
$\tilde{\gamma}_{r1(2)}$ are determined by expressions
\begin{equation}
\label{5}
\tilde{\gamma}_{r1(2)}=\gamma_{r1(2)}\,|R(\omega_\ell\nu/c)|^2,
\end{equation}
where
\begin{equation}
\label{6} R(\kappa)=\int_{-d/2}^{d/2}\,dz\,exp(-i\kappa z)\Phi(z),
\end{equation}
$\nu$ is the light refraction factor, which for simplicity is
considered as identical for a QW and barrier, $\Phi(z)$ is the
wave function of excitation at $z_e=z_h=z$, where $z_{e(h)}$ is
the electron (hole) coordinate. The factor $|R(\omega_\ell\nu/c)
|^2$ determines the dependence of $\tilde{\gamma} {r1l(2)}$ on the
QW width. The radiative energy shifts $\Delta_{1(2)}$ also depend
on QW width $d$ and aspires to zero at $\kappa d\to 0$.

In a limiting case $\tilde{\gamma}_{1(2)}\ll\gamma_{1(2)},~~
\Delta_{1(2)}\ll\gamma_{1(2)}$ the perturbation theory on
interaction of light with electronic system is applicable. Thus,
${\cal R}\ll 1, {\cal A}\ll 1$ and ${\cal R}\ll{\cal A}$. If
radiative and non-radiative damping are comparable, ${\cal R}$ and
${\cal A}$ achieve as much as possible allowable values about
unit. Most interesting results turn out under conditions
\begin{equation}
\label{7} \gamma_{rl(2)}\gg\gamma_{1(2)},\quad \Delta_{1(2)
}\gg\gamma_{1(2)}.
\end{equation}
If $\gamma_l=\gamma_2=0$, light absorption ${\cal A}=O$, and for
light reflection ${\cal R}$ from (2) we have
\begin{equation}
\label{8} {\cal
R}={[(\tilde{\gamma}_{r1}+\tilde{\gamma}_{r2})/2]^2
(\omega_\ell-\Omega_0)\over (\omega_\ell-\omega_{d1})^2
(\omega_\ell-\omega_{d2})^2+[(\tilde{\gamma}_{r1}+\tilde{\gamma_{r2}})/2]^2
(\omega_\ell-\Omega_0)^2}
\end{equation}
where
$$\Omega_0=(\omega_1\tilde{\gamma}_{r2}+\omega_2\tilde{\gamma}_{r1})
/(\tilde{\gamma}_{r1}+\tilde{\gamma}_{r2}),$$
$$\omega_{d1(2)}=(1/2)[\omega_1+\Delta_1+\omega_2+\Delta_2 $$$$\pm
\sqrt{(\omega_1+\Delta_1-\omega_2-\Delta_2)^2+4\Delta_1\Delta_2}].$$
It follows from (8) that at $\omega\ell=\Omega 0$ ~~ ${\cal R}=O$,
i. e. there is a total transmission point of light through a QW.
At $\omega_\ell=\omega_{dl}$ or $\omega_\ell=\omega_{d2} $
~~${\cal R}=l$, i. e. light is totally reflected. The position of
points of total reflection depends on the QW width $d$.

\begin{figure}
\epsfxsize=100mm \centerline{\epsffile{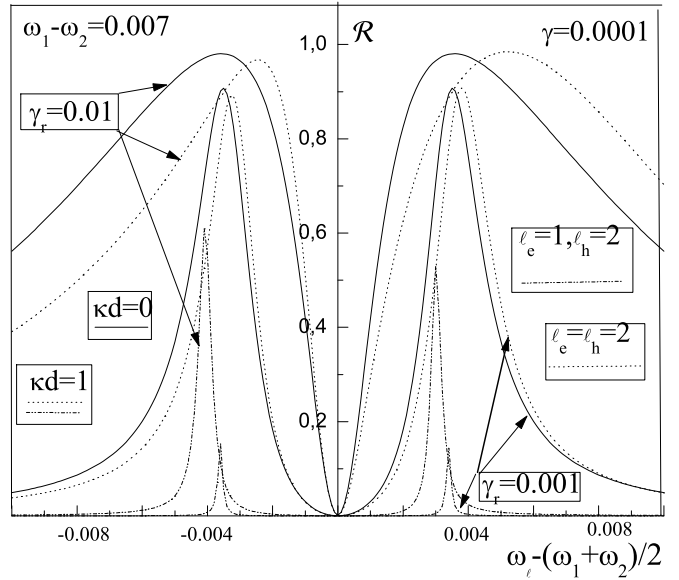}}
\caption[*]{Dimensionless  reflection ${ \cal R} $  as function of
light frequency $ \omega_\ell $ in case of two excitation energy
levels
 in a wide QW under condition $ \gamma_r\gg\gamma $.
Continuous and dashed lines are the permitted transitions,
dot-and-dash lines are the forbidden transitions. $ \gamma_r
(\gamma) $ is the radiative (non- radiative) broadening of an
exited state.}
 \label{Fig.1}
\end{figure}
\begin{figure}
\epsfxsize=100mm \centerline{\epsffile{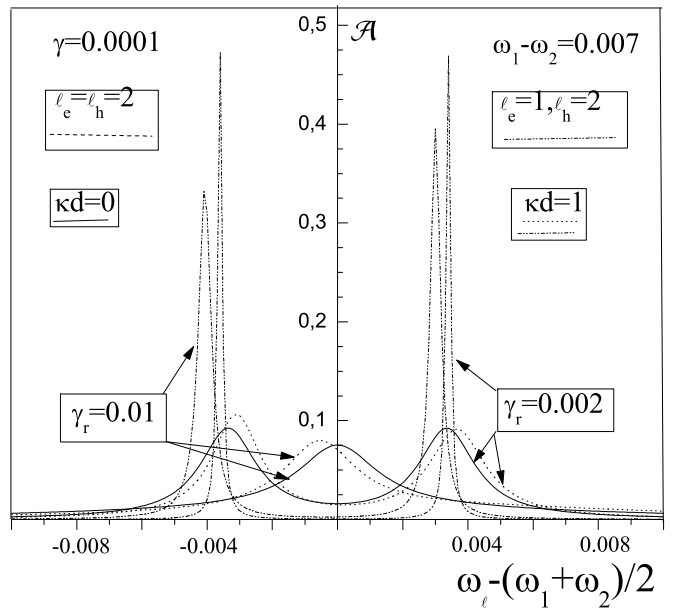}}
\caption[*]{Same, that in Fig.1 for dimensionless  absorption
${\cal A} .$ }
 \label{Fig.2}
\end{figure}
Represented in Fig. 1 and 2 frequency dependencies ${\cal R}$ and
${\cal A}$ correspond to inequalities (7). These figures may
concern and to weak polarons, when $\Delta E_{weak}$ and
$\tilde{\gamma}_{rl(2)}$ may be comparable.

So, it is possible to make following conclusions. First, it is
shown that if a light wave length is comparable to a QW width,
height and form of peaks on curves ${\cal R} (\omega_\ell) $ and
${\cal A} (\omega_\ell) $ depend on a QW width.

Secondly, it is established that in case of a wide QW the
interaction of light with excitations, which are characterized by
different numbers $\ell_e\neq\ell_h$ of size of quantization
numbers of electrons and holes  appears. Dependence of ${\cal R}$
and ${\cal A}$ on a QW width $d$ is various for cases
$\ell_e=\ell_h$ and $\ell_e\neq\ell_h$.

Thirdly, it is shown that for wide QWs is kept original behavior
of ${\cal R}$ depending on frequency $\omega_\ell$ (i. e.
existence of two points of total reflection and one point of total
transmission if the non-radiative damping is much less than
radiative one.

\end{document}